\documentclass[11pt,twoside]{article} 
\usepackage{cspm-asp2006}
\usepackage{epsfig,graphicx,natbib,url}
\usepackage{lscape} 
\begin{document}
\setcounter{page}{247}

\markboth{Labrosse et al.}{\ion{O}{vi} and H$_2$ Lines in Sunspots}
\title{O VI and H$_2$ Lines in Sunspots}
\author{Nicolas Labrosse$^1$, Huw Morgan$^2$, Shadia Rifai Habbal$^2$\\ and Daniel Brown$^1$}
\affil{$^1$Institute of Mathematical and Physical Sciences, Aberystwyth, UK\\
       $^2$Institute for Astronomy, University of Hawaii, USA}

\begin{abstract}
Sunspots are locations on the Sun where unique atmospheric conditions
prevail. In particular, the very low temperatures found above sunspots
allow the emission of H$_2$ lines. In this study we are interested in
the radiation emitted by sunspots in the \ion{O}{vi} lines at 1031.96
\AA\ and 1037.60\,\AA. We use SOHO/SUMER observations of a sunspot
performed in March 1999 and investigate the interaction between the
\ion{O}{vi} lines and a H$_2$ line at 1031.87\,\AA\ found in the Werner
band. The unique features of sunspots atmospheres may very well have
important implications regarding the illumination of coronal O$^{+5}$
ions in the low corona, affecting our interpretation of Doppler
dimming diagnostics.
\end{abstract}

\section{Introduction}

The \ion{O}{vi} lines at 1031.96\,\AA\ and 1037.60\,\AA\ are far more
intense in sunspots than in the Quiet Sun
(\cite{nl-foukaletal74,nl-sumdiscatlas}). The contribution of sunspots
to the total integrated disk spectrum seen by O$^{+5}$ ions in the
corona is therefore non negligible. This has important consequences on
the interpretation of observed \ion{O}{vi} lines in the corona
(\cite{nl-huw05}). Our goal is to understand the formation mechanisms
of these lines in sunspots. We present an analysis of SUMER
observations of a sunspot from March 18, 1999. We discuss the
interaction between the \ion{O}{vi} lines and the emission line of
molecular hydrogen H$_2$ at 1031.87\,\AA.

\section{Observations}

\begin{figure}
  \centering
  \plottwo{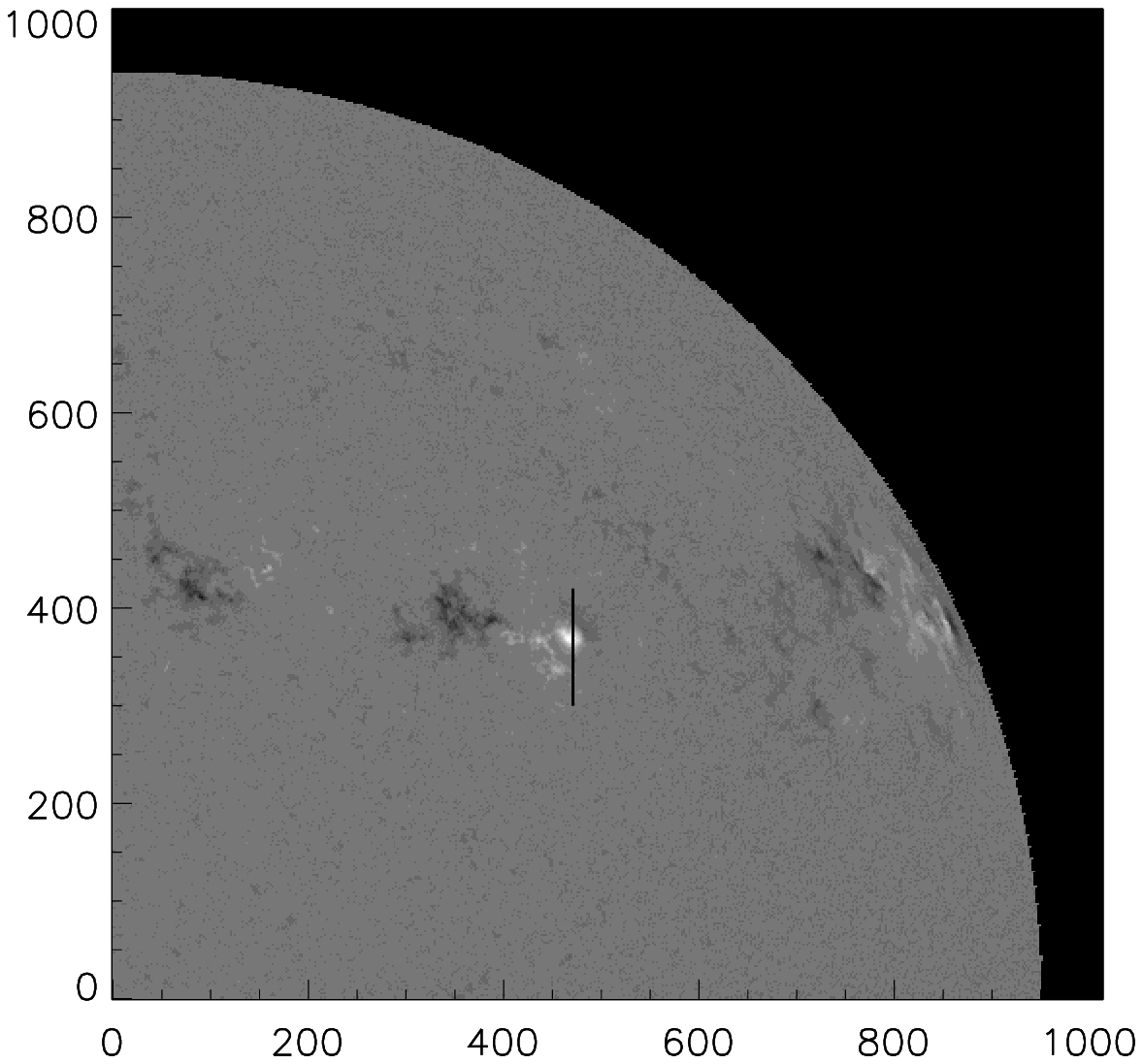}{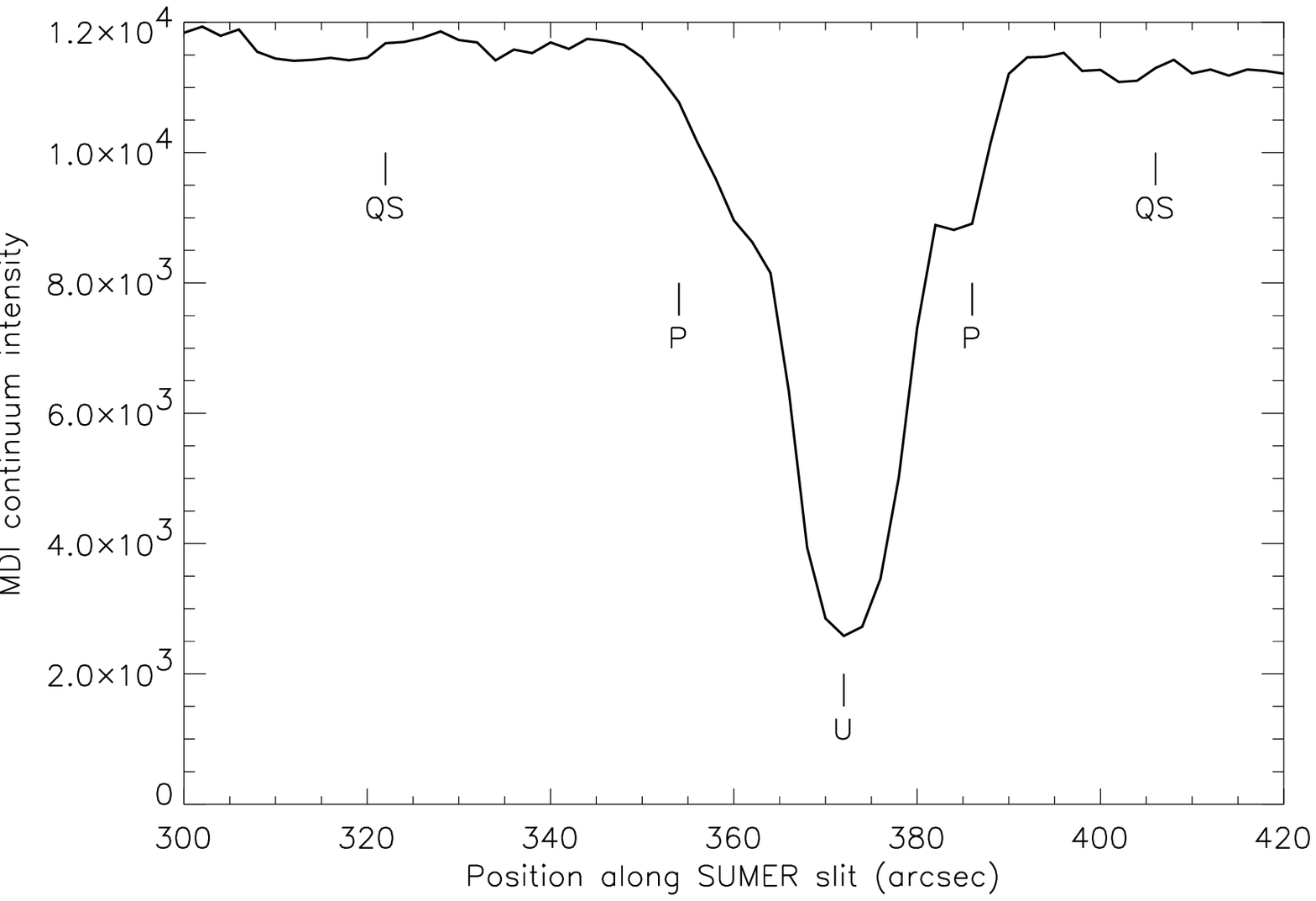}
  \caption[]{\label{nl-fig:sunspot context}
  {\em Left:} MDI image of sunspot with SUMER slit. {\em Right:} MDI continuum intensity along SUMER slit.}
\end{figure}

We use the reference spectrum obtained by the SUMER spectrometer
onboard the SOHO spacecraft on March 18, 1999
(\cite{nl-schuetal99}). These observations were made with the detector
B and lasted for 7 hours. The spectrometer slit
(0.3\arcsec$\times$120\arcsec) was kept at a fixed position centered
on the sunspot {(Fig.~\ref{nl-fig:sunspot context}, left)} and
compensation for the solar rotation was turned on.  From the MDI
continuum intensity we estimate the location of the sunspot umbra to
be between 364\arcsec\ and 382\arcsec\ on the SUMER slit
({Fig.~\ref{nl-fig:sunspot context}, right}). The surrounding penumbra
is between 344\arcsec\ and 364\arcsec, and between 382\arcsec\ and
392\arcsec. The upper and lower parts of the slit are on the quiet
sun.  These features will now be denoted U, P, and QS, respectively.

\begin{figure}
    \plottwo{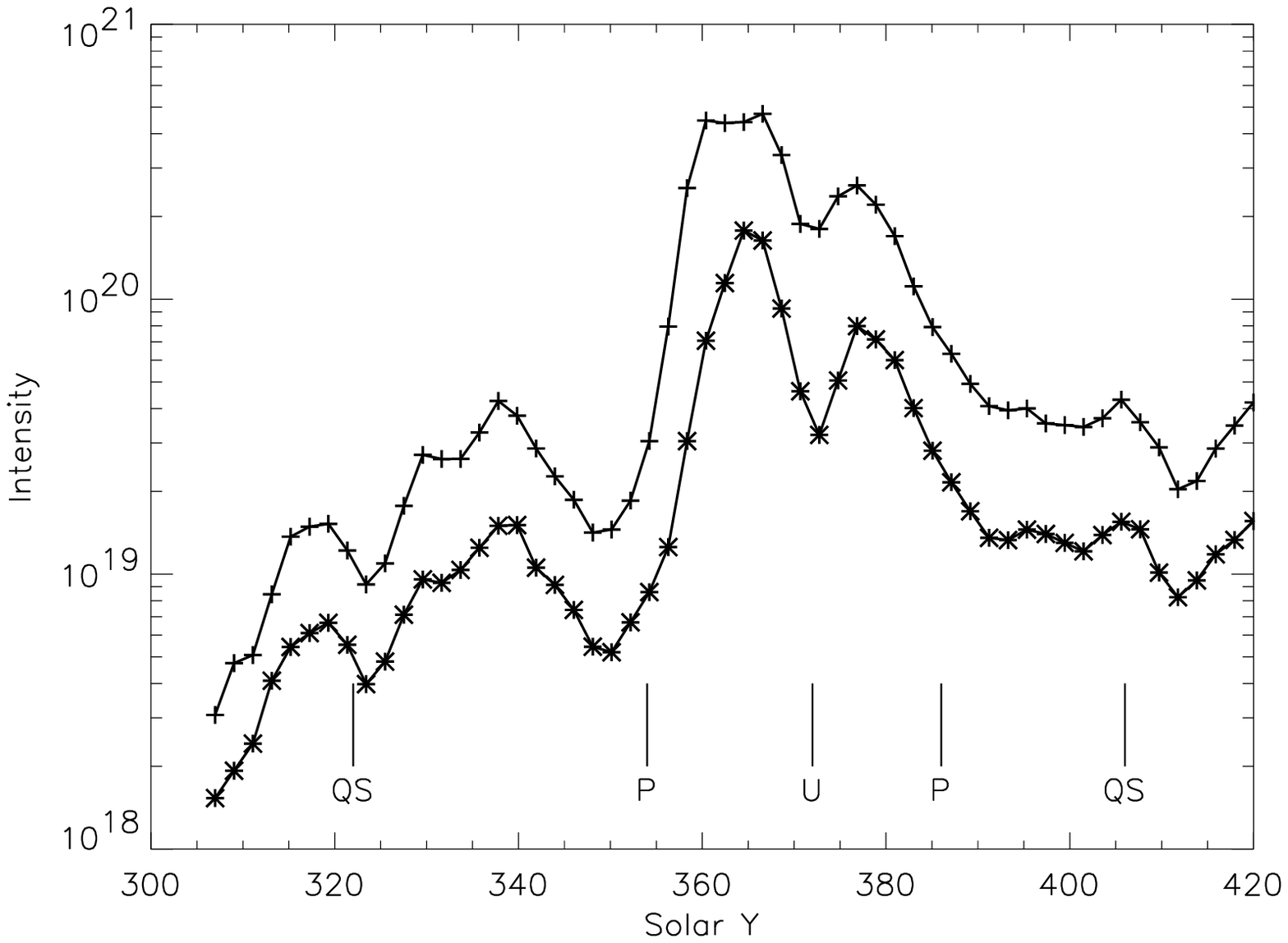}{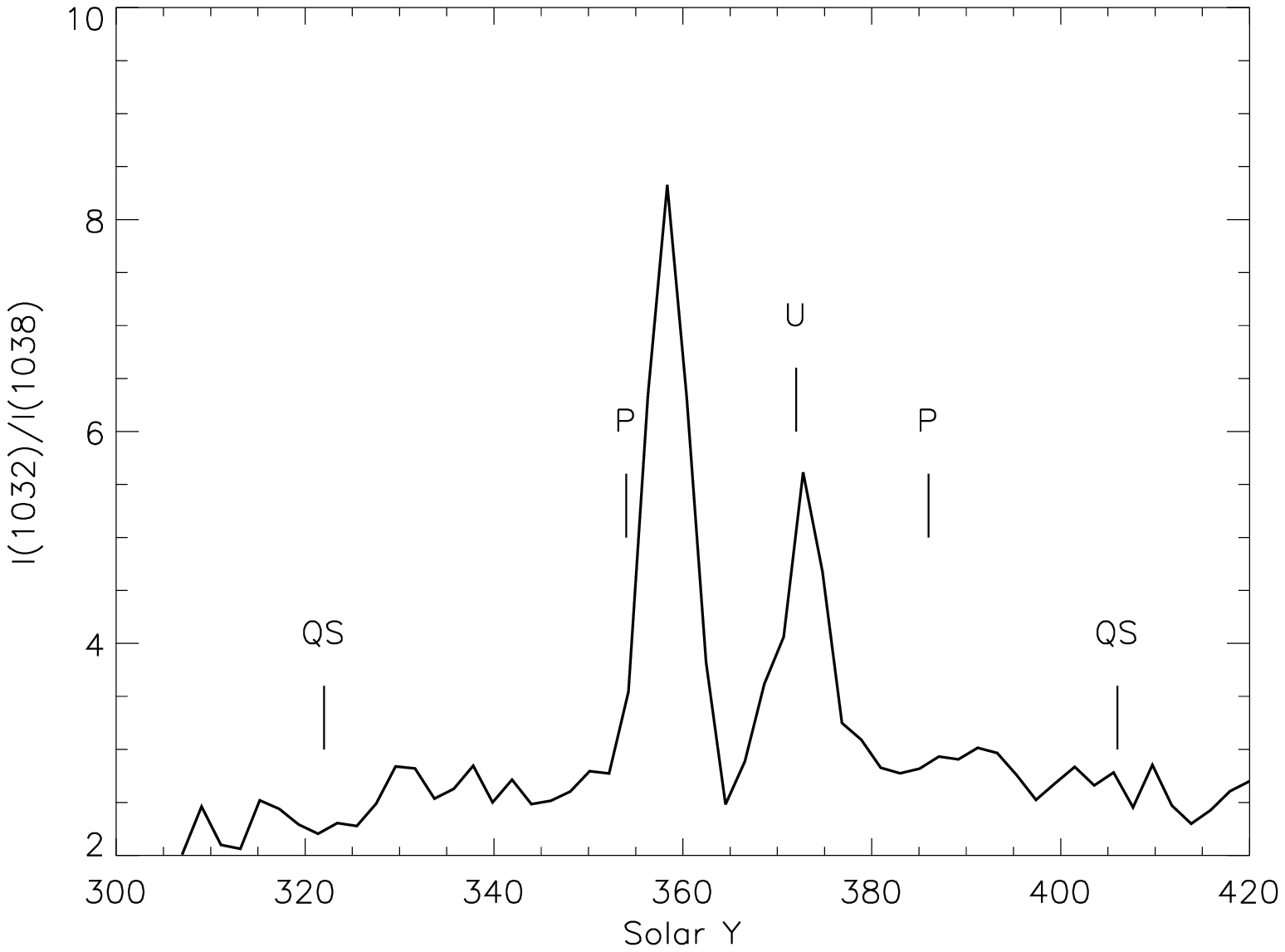}
    \caption[]{{\em Left:} Integrated intensity of \ion{O}{vi} 1032\,\AA\ (+) and 1038\,\AA\ (*) lines. The line intensities are enhanced in the sunspot region compared to the quiet sun, particularly in the penumbra/umbra interface.
    {\em Right:} Intensity ratio of \ion{O}{vi} lines along SUMER slit.}
    \label{nl-fig:ovi}
  \end{figure}

The two panels of {Fig.~\ref{nl-fig:ovi}} show how the intensities of
the \ion{O}{vi} lines are affected by the different physical
conditions in the quiet sun, in the penumbra, and in the umbra. The
integrated intensities have been determined by a simple integration
over the line profiles.  The intensity profile of the \ion{O}{vi}
lines is similar to what has been observed by, e.g.,
\citet{nl-foukaletal74}.

An interesting and important feature is that the ratio between the two
lines significantly deviates from its average value. While it is
around 2 on QS, it reaches values greater than 8 in the southern
umbra/penumbra interface. This implies that the coronal O$^{5+}$ ions,
which are sensitive to the disk radiation, are illuminated by a
radiation field which may very well be different from the averaged one
(QS) that is usually assumed. Doppler dimming diagnostic techniques
using the \ion{O}{vi} doublet line ratios, are sensitive to the value
of the ratio on the disk. {Adopting a value of 2 for the line ratio
and a homogeneous disk could lead to spurious results of ion speeds in
the corona} considering the impact that sunspots can have on the disk
radiation field.  This is due to the presence of H$_2$ emission at
1031.87\,\AA\ (\cite{nl-bartoeetal79}) in the 1--1 Werner band. This
molecular line is excited by resonance fluorescence from the
\ion{O}{vi} line.

\section{Line Formation}

\begin{figure}
    \plottwo{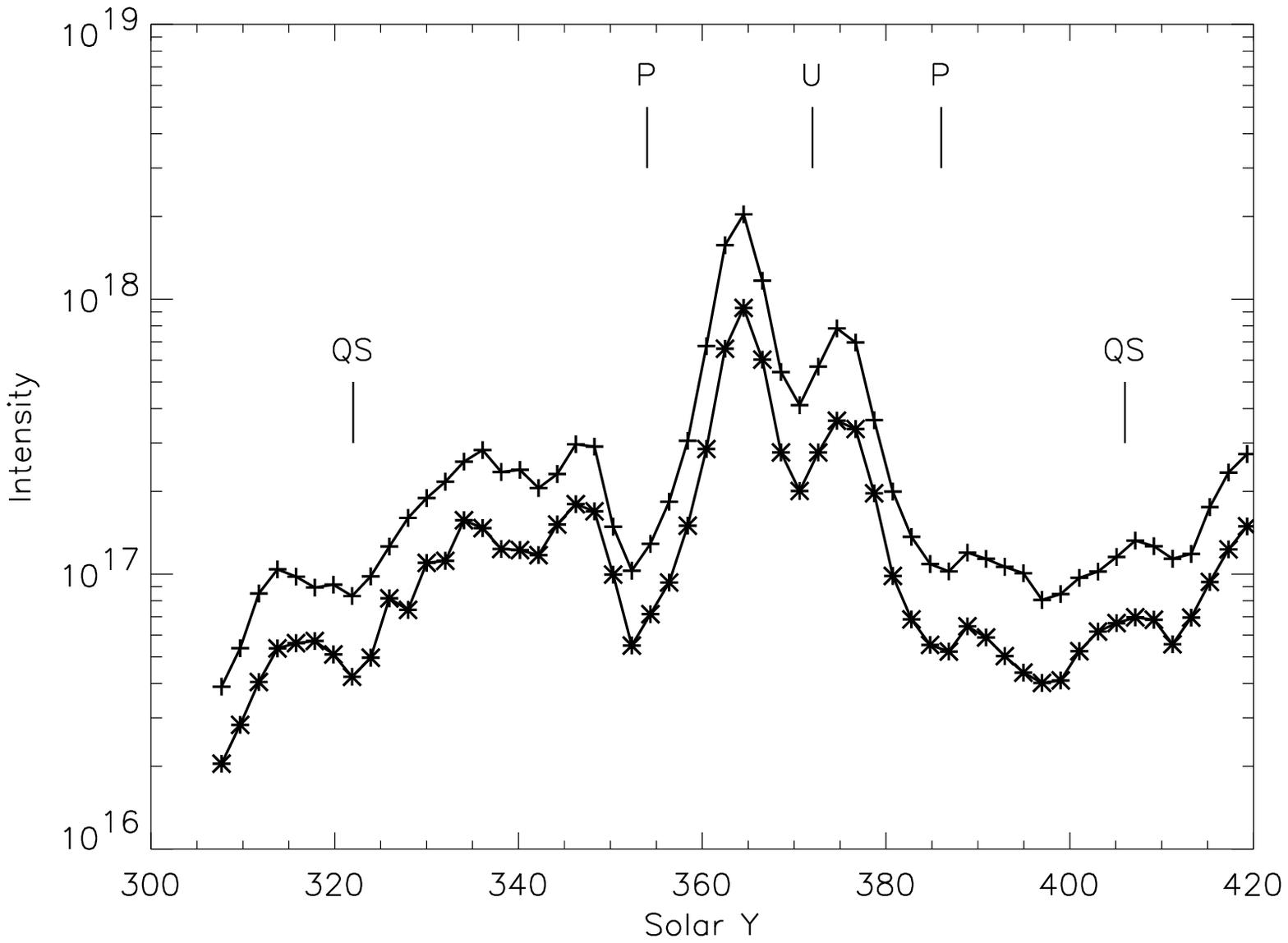}{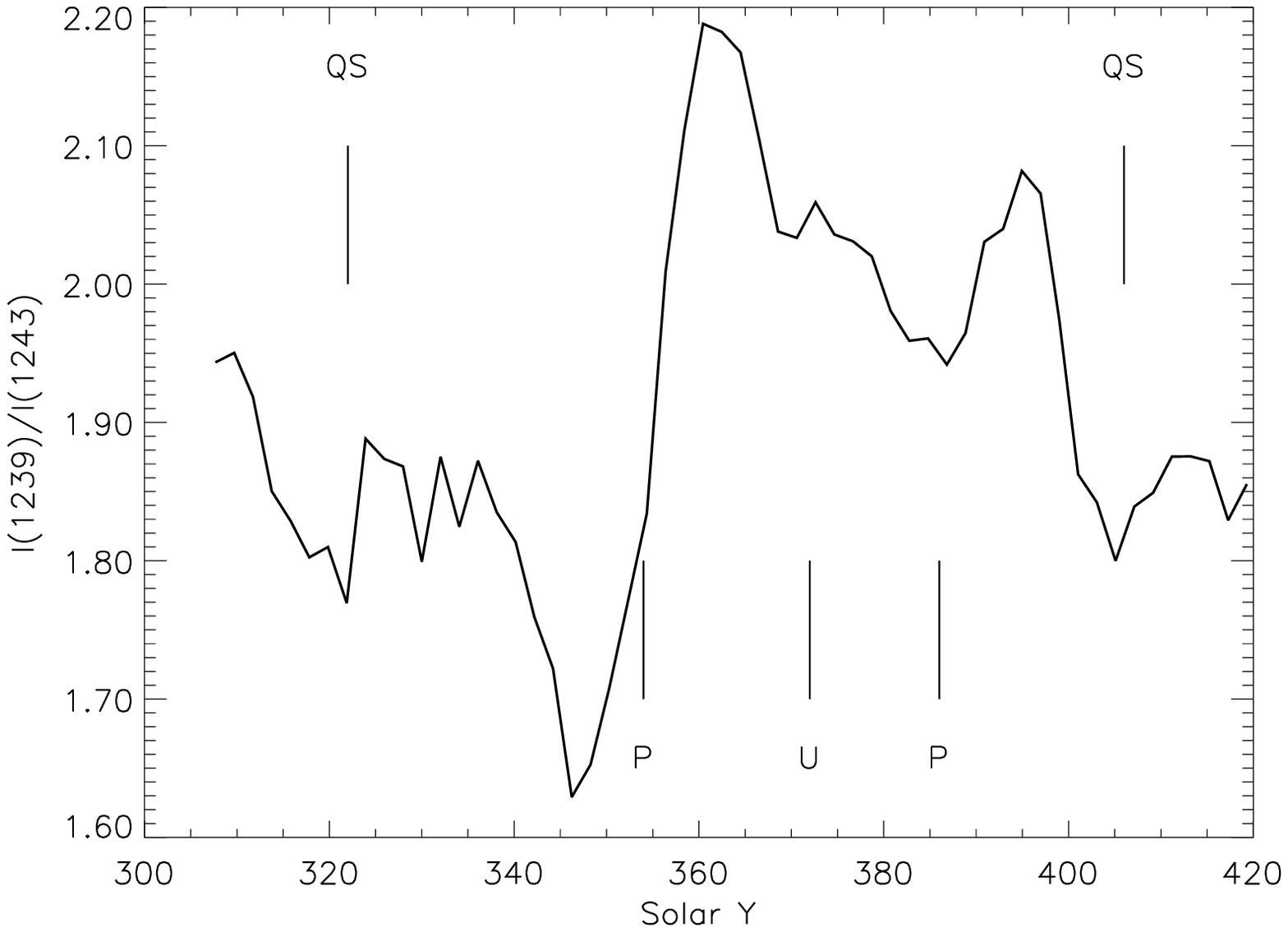}
    \caption[]{{\em Left:} Integrated intensities of \ion{N}{v} 1239 (+) and 1243 (*) lines. {\em Right:} Intensity ratio of \ion{N}{v} lines along SUMER slit.}
    \label{nl-fig:nv}
  \end{figure}

The two panels in {Fig.~\ref{nl-fig:nv}} present the integrated
intensities and intensity ratio of the two \ion{N}{v} lines at 1239
and 1243\,\AA\ from the same data set as \ion{O}{vi}. The \ion{N}{v}
and \ion{O}{vi} doublets are in the lithium-like sequence.  The
\ion{N}{v} doublet is very similar to \ion{O}{vi} (same collisional
excitation coefficient ratio and similar temperature of formation), so
the \ion{N}{v} 1239/1243 intensity ratio should be close to 2 at all
temperatures, even accross the sunspot. This proves the special case
of \ion{O}{vi} due to its interaction with the H$_2$ lines in the Q3
band.

Furthermore, the SUMER data shows that the \ion{C}{ii} lines at 1036.3
and 1037\,\AA\ are fainter in the sunspot region. This is expected as
the lines of first ions are reduced in intensity over sunspots
(\cite{nl-bartoeetal79}).  {This may in turn have implications for the
pumping of the coronal \ion{O}{vi} 1037\,\AA\ line at speeds above
$\sim$100~km/s.}

  \begin{figure}
    \centering
    \includegraphics[width=0.8\textwidth]{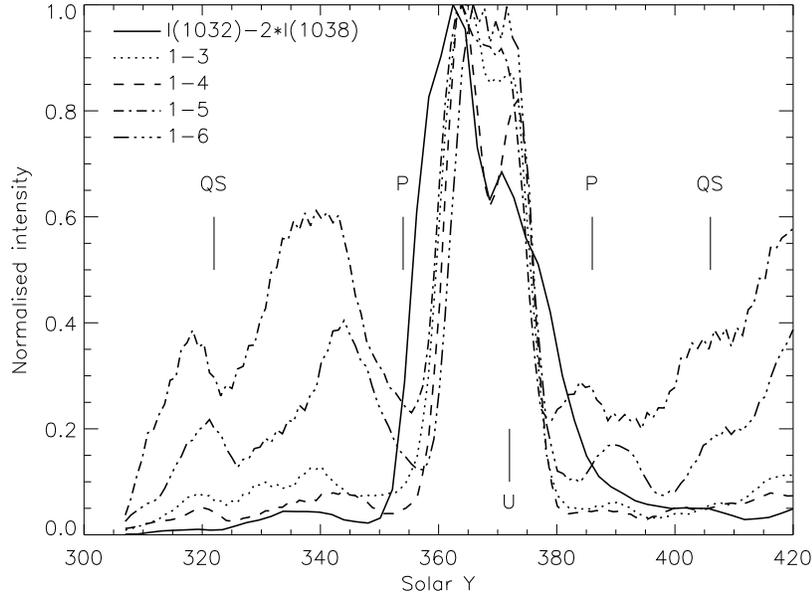}
    \caption[]{Normalized intensity profiles of four unblended H$_{2}$ Werner band lines across the sunspot. The solid line shows the difference I(\ion{O}{vi} 1032)-2$\times$I(\ion{O}{vi} 1038). The enhancement of intensity at the sides of the sunspot is interesting,  and is possible evidence of ambipolar diffusion of H$_2$ from the umbra, that is, H$_2$ and other neutrals can diffuse across the near-vertical magnetic field of the sunspot umbra (\cite{nl-km06}).}
    \label{nl-fig:ovidiff}
  \end{figure}

Figure~\ref{nl-fig:ovidiff} helps us to quantify the amount of H$_2$ molecular emission present in the SUMER sunspot observations by assuming that the integrated intensity at 1032\,\AA\ is normally the double of the intensity at 1038\,\AA. This figure also presents the normalized intensity profiles of four H$_2$ Werner band lines across the sunspot. These lines are not blended with other UV lines, therefore they predict what the 1031.87\,\AA\ 1--1 line would look like in the absence of interaction with \ion{O}{vi}. The relative intensity of all these lines agree closely with the calculated emission transition probabilities (see \cite{nl-schuetal99}). 
We can see that H$_2$ and O$^{5+}$ interaction is significant in the sunspot area only.

\section{Conclusion}

The sunspot's molecular H$_2$ line interacting with the \ion{O}{vi}
lines is one of many in the H$_2$ Werner band. Theory gives relative
intensities for each line, and this theory has been tested by
\citet{nl-schuetal99} on the SUMER sunspot data. This means that we
can know very closely the H$_2$ emission in the 1--1 band at
1031.87\,\AA, which is the one that pumps the \ion{O}{vi} 1031.9\,\AA,
just by looking at a clean line (e.g., the 1--3 band at 1119\,\AA) and
multiplying by the theoretical intensity ratio between the two given
in \citet{nl-bartoeetal79}.

In a future work we will model a slab of oxygen ions in the transition
region above a sunspot and calculate the 1032 and 1038 line
intensities emitted by the slab purely by collisions. Then, using a
radiative transfer code, we will add the H$_2$ line intensity at the
bottom of the slab, and study the pumping of \ion{O}{vi} 1032\,\AA\ that
ensues. 

\acknowledgements N.\,L. acknowledges financial support from the
organisers of the Coimbra Solar Physics Meeting, the University of
Wales through the Gooding Fund, and PPARC through grant
PPA/G/O/2003/00017.

\end{document}